# Temperature Dependent Performances of Superconducting Nanowire Single-Photon Detectors in an Ultralow-Temperature Region

**Taro Yamashita,[*] Shigehito Miki, Wei Qiu, Mikio Fujiwara,[1] Masahide Sasaki,[1] and Zhen Wang**

*Kobe Advanced ICT Research Center, National Institute of Information and Communications Technology, 588-2 Iwaoka, Iwaoka-cho, Nishi-ku, Kobe, Hyogo 651-2492, Japan*

*[1]National Institute of Information and Communications Technology, 4-2-1 Nukui-Kitamachi, Koganei, Tokyo 184-8795, Japan*

*[*]E-mail address: taro@nict.go.jp*

We present the performances of a superconducting nanowire single-photon detector (SNSPD) in an ultralow-temperature region from 16 mK to 4 K. The system detection efficiency of the SNSPD showed saturation in the bias-current and temperature dependences, and reached the considerably high value of 15% for 100 Hz dark count rate below 1.4 K at a wavelength of 1550 nm even without an optical cavity structure. We found that the dark count exists even at 16 mK and black body radiation becomes its dominant origin in the low temperatures for fiber-coupled devices.

In the field of quantum information processing and sensing, superconducting nanowire single-photon detectors (SNSPDs) are promising components because they have the sensitivity covering from visible to mid-infrared wavelengths with low dark count rate (DCR).[1] They can operate at sub-GHz with excellent time resolution, and have actively been employed for many applications such as quantum key distribution (QKD), single-photon source characterization, and photon-counting optical communications.[2-7] In these applications, however, a further improvement in the performance is highly desirable, and significant efforts are being made to increase the system detection efficiency (DE) and reduce the DCR. Fundamental understanding of the devices and technical development of the systems are both crucial to improve the overall performances.

From a practical point of view, a fiber-coupled technique is known to be suitable for a practical SNSPD system because of its efficient optical coupling between incident photons and nanowire area.[6,8-12] In fiber-coupled SNSPD systems, relatively large (e.g., $20 \times 20\ \mu m^2$) devices have often been used. The system DE of the devices has been reported as 1 – 3% at around 3 K and 100 Hz DCR for a wavelength of 1550 nm.[8-10] Although the quantum efficiency of 17% at 2 K has been reported in the experimental setup using an input window,[13] the improvement of the system DE in the fiber-coupled system is highly required for a practical SNSPD system utilized in the applications.[2-7] Furthermore, the DE of the large devices has not reached the intrinsic one as achieved in small devices.[14]

It is well known that the device DE and DCR directly depend on the dynamics of the resistive-state formation process in the nanowire. Although several theoretical models have been proposed for explaining it (e.g., unbinding of vortex-antivortex pairs),[15-19] the physical mechanism has not been clearly identified yet. One of the effective ways is to investigate the properties of SNSPDs at ultralow temperatures because the intrinsic physical characteristics and the ultimate potentials appear most in the vicinity of the absolute zero temperature. At higher temperatures, they are often concealed due to thermal fluctuations. Systematic measurements of the temperature dependences of the system DE and DCR over a wide range of temperatures would also be useful for the improvement of the SNSPD performances as well as the clarification of the temperature-dependent physical characteristics.

In this letter, we report the first measurements of the temperature-dependent performances of a fiber-coupled SNSPD in an ultralow-temperature region from 16 mK to 4 K. We describe in detail the bias-current and temperature dependences of the system DE and DCR of the fiber-coupled SNSPD system, and discuss the origin of the dark count.

The SNSPD device used in the present work was fabricated as follows. NbN films with a thickness of 4 nm were grown on a single-crystal MgO substrate by reactive dc magnetron sputtering. The NbN thin film was formed up to a 100-nm-wide meander line covering an area of $20 \times 20\ \mu m^2$ and with a filling factor of 62.5%. The superconducting critical temperature $T_C$ of the device was 8.8 K, and the critical current density $J_C$ was $3.1 \times 10^{10}$ A/m$^2$ at 3.0 K. The fabrication process and dc characteristics of SNSPDs have been described in detail elsewhere.[20]

Figure 1 shows a schematic diagram of our ultralow-temperature measurement system. The nanowire device was placed in a fiber-coupled package.[6,9] The fiber-coupled package was mounted on an Au-plated oxygen-free copper platform thermally connected to the mixing chamber of the dilution refrigerator with a base temperature of 11 mK. The sample temperature was indicted by a $RuO_x$ resistance thermometer (calibrated with a nuclear orientation thermometer down to 11 mK) located at the bottom of mixing chamber. The dilution unit is inside a fridge dewar mounted on a vibration-free stage. Double μ-metal cylinders surround the fridge dewar and form magnetic shields. The temperature was set using a Proportional-Integral-Derivative controller and all data were recorded when the temperature was stable. A single-mode (SM) optical fiber and a 50 ohm coaxial cable were connected from the top of the fridge to the SNSPD package. The device was current-biased via the dc arm of the bias tee, and the output signal passing through the ac arm of the bias tee and two low-noise amplifiers was observed using the pulse counter. A tunable-wavelength (1465 – 1575 nm) continuous-wave laser diode was used as the photon source. The input light was heavily attenuated so that the photon flux at the input connector of the cryostat was $10^6$ photons/s. A polarization controller was inserted in front of the optical input of the refrigerator to obtain the maximum system DE. The system DE was defined as the output count rate divided by the input rate of photon flux to the system.

Figure 2(a) shows the system DE and DCR for the photon source with 1550 nm wavelength as functions of the normalized bias current $I_{\mathrm{bias}}/I_C$ at 16 mK, 2.0, 3.0, and 4.0 K. The system DE increased drastically with a decrease in the operation temperature. At the temperature of 16 mK, we first observed the highest DE value of 15% at a DCR of 100 Hz and a saturation region indicating that the system DE almost reached the intrinsic DE despite its large nanowire area (20 × 20 $\mu m^2$). Until now, an important factor that limits the device DE was the constrictions in the nanowires.[14)] It is, however, likely that the drastic improvement of the DE observed here is not solely due to the constrictions but also due to some other and maybe fundamental mechanisms in the nanowire.

Meanwhile, the DCR increased with the bias current at all temperatures. Two different slopes can be clearly seen in the bias-current dependence of the DCR as the operation temperature decreases and seem to originate from two different causes. The point at which the slope changes depends on the temperature, e.g., $I_{\mathrm{bias}} \sim 0.96\ I_C$ at 16 mK. In the bias-current region below the point where the slope changes, the DCR increased with a decrease in the temperature and the bias-current dependence was not linear but followed a curve similar to the one depicting the bias-current dependence of the DE. This behavior has been observed for the first time. Since we darkened the optical fiber port perfectly to eliminate any stray light entering the device, the most plausible cause is the blackbody radiation invading through the optical fiber connected to the optical input port at room temperature.[21)] As shown in Fig. 2(b), we measured the DCR at 3 K for a device mounted in the package with and without an optical fiber as a function of the normalized bias current, and confirmed that there is no dark count in the low bias region for the device without a fiber. As a result of the improvement of the DE at ultralow temperature and good optical coupling to the incident light, the sensitivity to the black body radiation increased. Although the DCR in this region was relatively low (with a maximum value of several tens of hertz at 16 mK), these dark counts would increase as the system DE improves further. It implies that additional countermeasures, such as narrow band filtering, should be employed for a further reduction of the DCR.

On the other hand, in the bias-current region above the point where the slope changes, the DCR showed steep slopes and decreased with a decrease in the temperature. It is obvious that the origin of the dark count in this region is related to the temperature, and it exists even at ultralow temperatures where thermal fluctuations are suppressed. The temperature dependence of the DCR in this region could offer valuable insights to identify the origins of intrinsic dark count events in the SNSPD by further analyses.

Figure 3 shows the system DEs as a function of the temperature at 10 and 100 Hz DCRs. The system DE increased with a decrease in the temperature, and reached a high value of 8% at 500 mK even for 10 Hz DCR. At 100 Hz DCR, the system DE saturated at temperatures below 1.4 K and showed a high value of 15%. This level of the DE is considerably larger than typical values (1-3%) of fiber-coupled SNSPDs without an optical cavity structure.[8-10] This remarkable improvement of the system DE has a great impact on a QKD experiment and leads to a higher key distribution rate.[2-5] We estimate that the saturation temperature would depend on the processing accuracy of the nanowire and the thin-film quality as well as the device design. The saturation of the DE at lower temperatures would be useful hints to uncover the mechanism of the resistive-state formation process of the nanowire.

In conclusion, we presented measurements and characterizations of the temperature dependences of the system DE and DCR for a fiber-coupled SNSPD in the ultralow temperatures from 16 mK to 4 K. Saturation in the DE versus the bias-current profile have been observed at 16 mK. A high system DE of 15% at a DCR of 100 Hz was achieved below 1.4 K, despite the large-area SNSPD device having no optical cavity structure. The system DE can be improved further by attaching an optical cavity to the device. We have also found that the dark count exists even at 16 mK and the dark count in the low bias region originates from black body radiation from room temperature. We believe that our results provide a considerable amount of information for clarifying the SNSPD mechanism, which can be helpful in achieving better performance.

## Figure captions

Fig. 1. Schematic diagram of an ultralow-temperature measurement system for the SNSPD.

Fig. 2. (a) System DE and DCR as functions of the bias current normalized by the critical current for incident photons with a wavelength of 1550 nm. The filled and open symbols indicate the system DE and DCR, respectively. (b) DCR as a function of the normalized bias current at 3 K. The cross and open circle indicate the DCR of a device mounted in the package with and without an optical fiber, respectively.

Fig. 3. Temperature dependence of the system DE for incident photons with a wavelength of 1550 nm. The circular and triangular symbols indicate the system DE at DCR values of 10 and 100 Hz, respectively.

# Figures

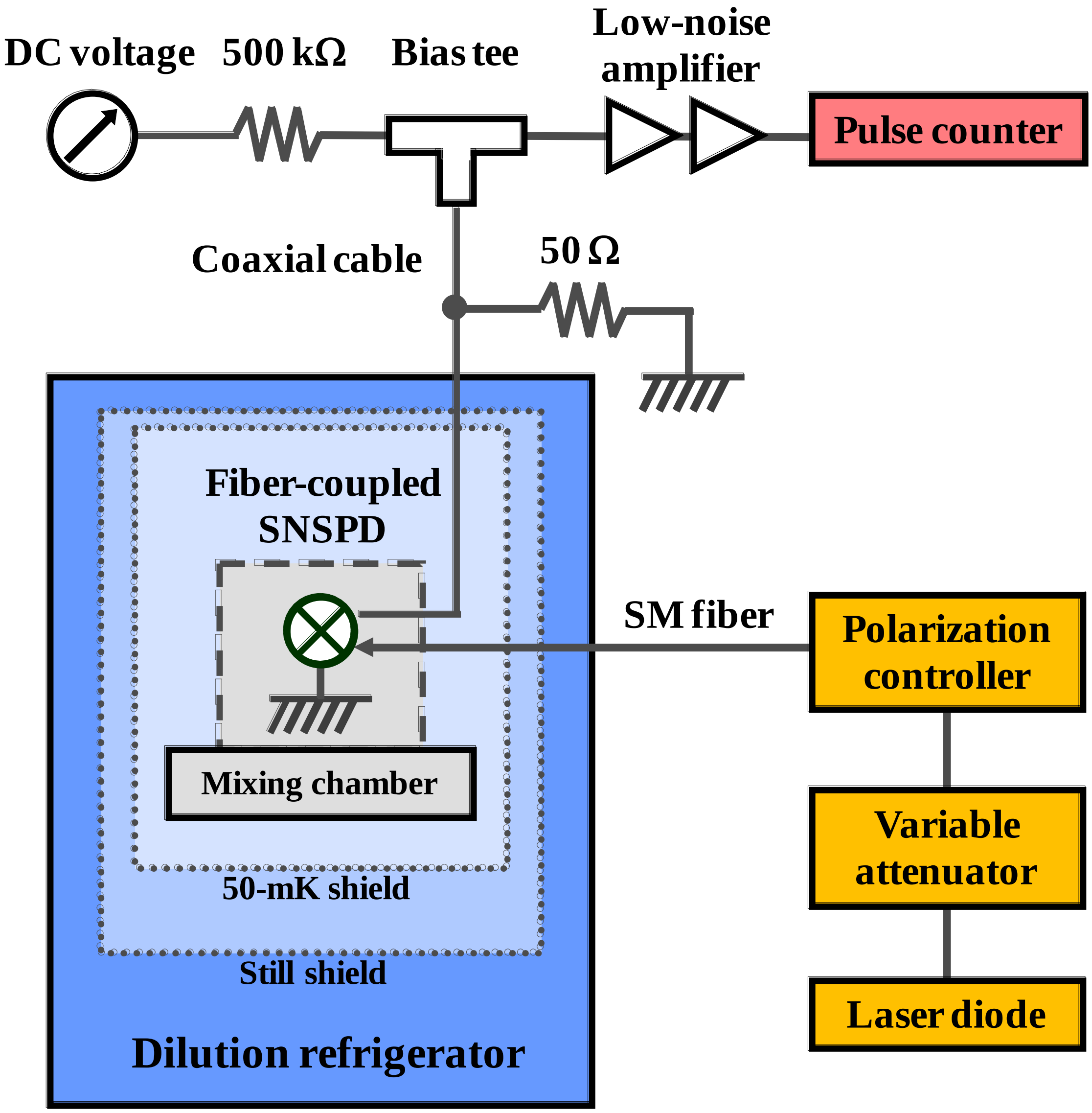

**Fig. 1**

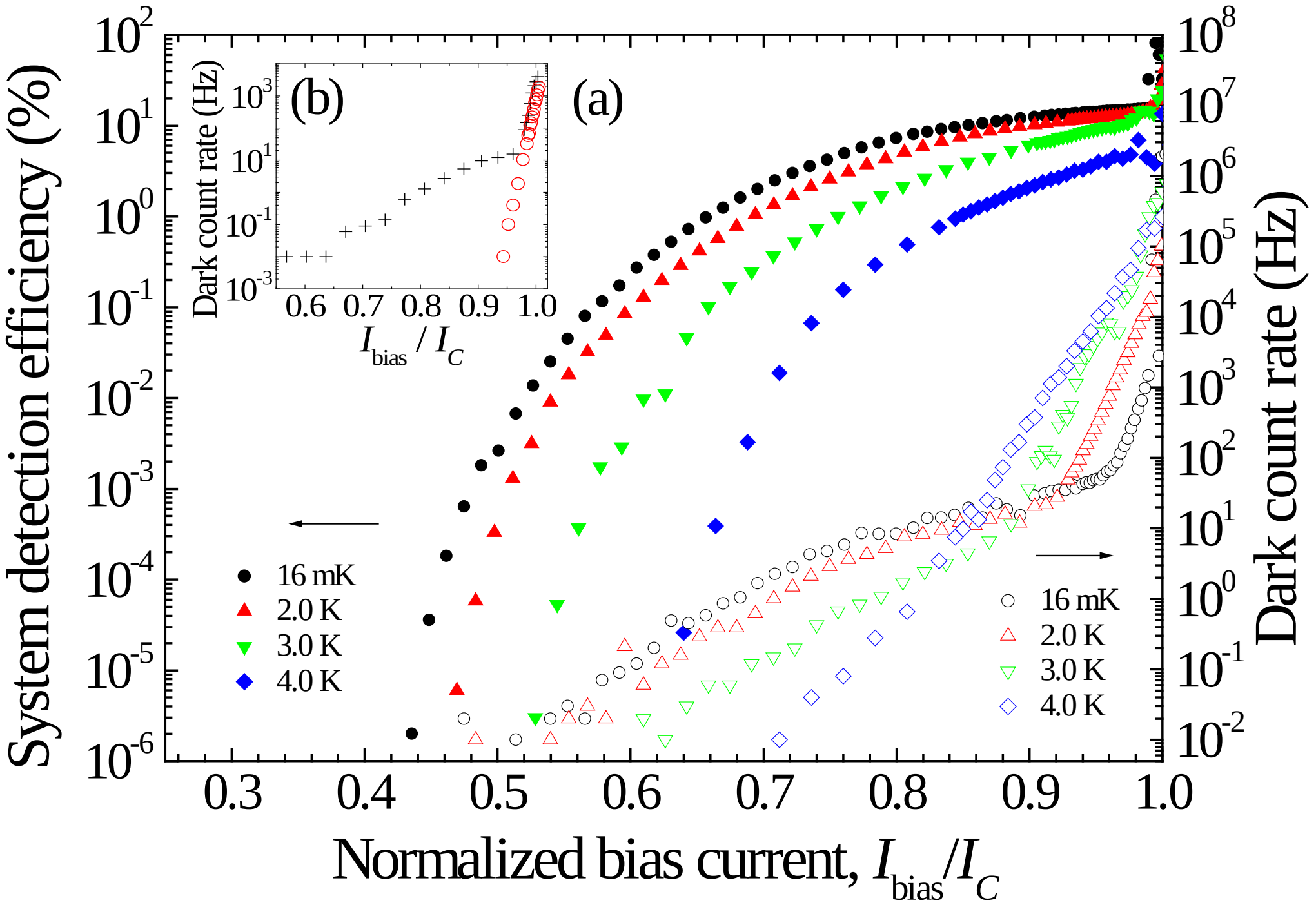
System detection efficiency (%)
Dark count rate (Hz)
Normalized bias current, $I_{bias}/I_C$
(a)
(b)
Dark count rate (Hz)
$I_{bias} / I_C$
16 mK
2.0 K
3.0 K
4.0 K
16 mK
2.0 K
3.0 K
4.0 K

**Fig. 2**

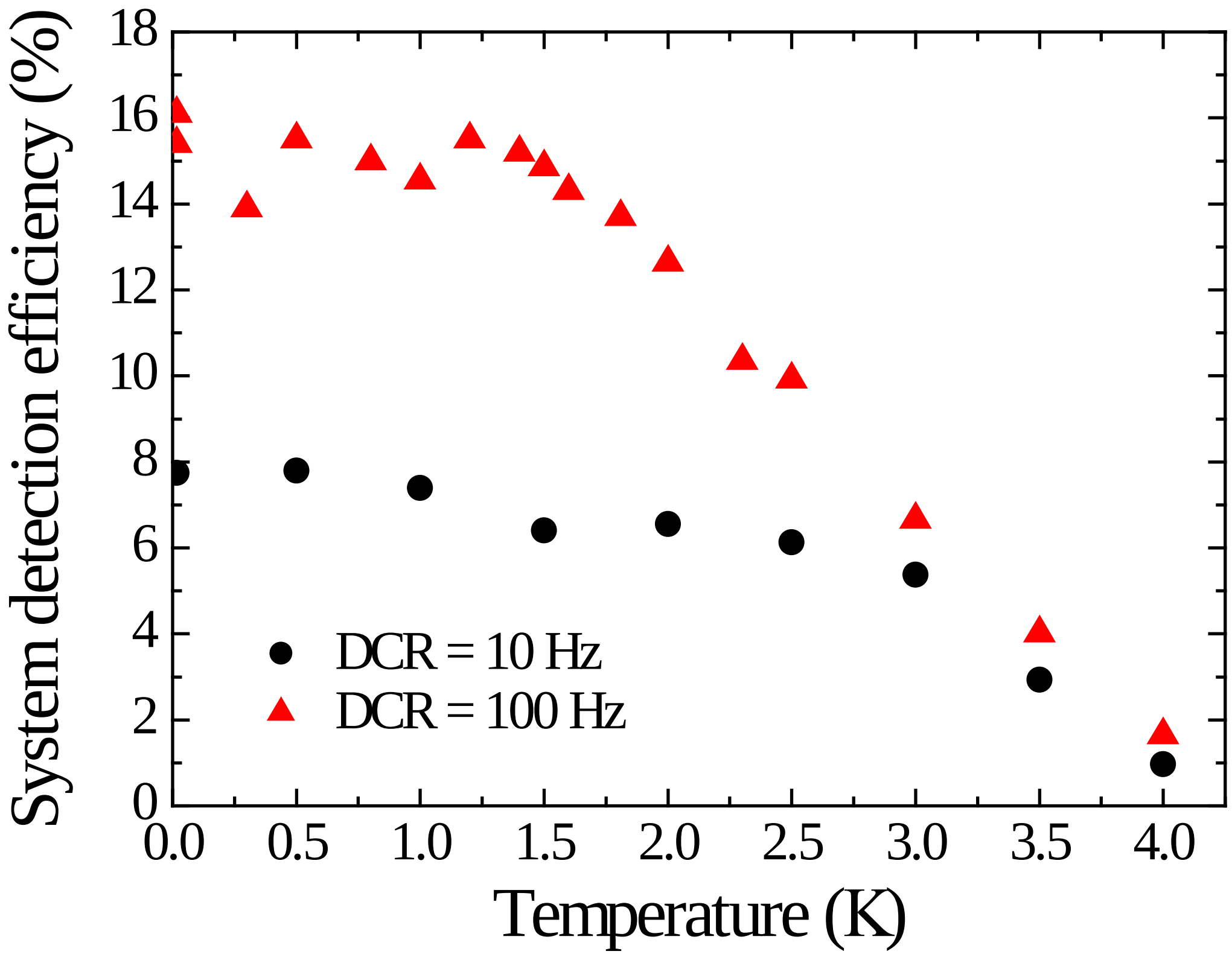

**Fig. 3**